\documentclass[aps,prl,twocolumn,superscriptaddress,longbibliography]{revtex4-2}
\usepackage{amssymb} 
\usepackage{graphicx,color} 
\usepackage{bbm} 
\usepackage{multirow}
\usepackage{amsmath}
\usepackage{mathrsfs} 
\usepackage{hyperref}
\usepackage[usenames,dvipsnames]{xcolor}
\usepackage{color}
\definecolor{darkgreen}{rgb}{0,0.6,0}
\definecolor{darkblue}{rgb}{0,0,0.6}
\definecolor{darkred}{rgb}{0.6,0,0}
\definecolor{darkpurple}{rgb}{0.5,0,0.5}

\hypersetup{
bookmarksopen=true,
pdftitle="",
pdfauthor="", 
pdftoolbar=false, 
pdfstartview={FitH},		
pdfmenubar=true,			
pdfhighlight=/O,			
colorlinks=true,			
urlcolor=darkblue,
citecolor=darkblue,		
linkcolor=darkblue}

\usepackage{tikz}
 
\usepackage{amsmath}
\usepackage{amsfonts}
\usepackage{cases}
\usepackage{xspace}
\usepackage{psfrag}
\usepackage{color}
\usepackage{xcolor}

\newcommand{\ave}[1]{\left\langle #1 \right\rangle}

\newcommand{\imag}{\mathring{\imath}}

\newcommand\ptwiddle[1]{\mathord{\mathop{#1}\limits^{\scriptscriptstyle(\sim)}}}

\newcommand{\norm}{\NC}

\newcommand{\plaind}{\mathrm{d}}
\newcommand{\dint}[1]{\mathchoice{\!\plaind#1\,}{\!\plaind#1\,}{\!\plaind#1\,}{\!\plaind#1\,}}

\newcommand{\Ndint}[1]{\ddintx{#1}{N}}
\newcommand{\ddintx}[2]{\mathchoice{\!\plaind^{#2}#1\,}{\!\plaind^{#2}#1\,}{\!\plaind^{#2}#1\,}{\!\plaind^{#2}#1\,}}

\newcommand{\dbar}{\plaind\mkern-6mu\mathchar'26}
\newcommand{\deltabar}{\delta\mkern-6mu\mathchar'26}
\newcommand{\dintbar}[1]{\mathchoice{\!\dbar#1\,}{\!\dbar#1\,}{\!\dbar#1\,}{\!\dbar#1\,}}

\usepackage{dsfont}

\newcommand{\canetset}[1]{{\mathchoice {\hbox{$\sf\textstyle #1\kern-0.4em #1$}}
{\hbox{$\sf\textstyle #1\kern-0.4em #1$}}
{\hbox{$\sf\scriptstyle #1\kern-0.3em #1$}}
{\hbox{$\sf\scriptscriptstyle #1\kern-0.2em #1$}}}}

\def\nbZ{{\mathchoice {\hbox{$\sf\textstyle Z\kern-0.4em Z$}}
{\hbox{$\sf\textstyle Z\kern-0.4em Z$}}
{\hbox{$\sf\scriptstyle Z\kern-0.3em Z$}}
{\hbox{$\sf\scriptscriptstyle Z\kern-0.2em Z$}}}}

\newcommand{\gpvec}[1]{\mathbf{#1}}

\newcommand{\yvec}{\gpvec{y}}

\let\AC\undefined
\newcommand{\AC}{\mathcal{A}}

\newcommand{\NC}{\mathcal{N}}

\newcommand{\TC}{\mathcal{T}}

\newcommand{\phitilde}{\tilde{\phi}}
\newcommand{\psitilde}{\tilde{\psi}}

\renewcommand{\exp}[1]{\mathchoice{\mathrm{e}^{#1}}{\operatorname{exp}\left(#1\right)}{\operatorname{exp}\left(#1\right)}{\operatorname{exp}\left(#1\right)}}

\newcommand{\elabel}[1]{\label{eq:#1}}
\newcommand{\eref}[1]{(\ref{eq:#1})}
\newcommand{\Eref}[1]{Eq.~(\ref{eq:#1})}
\newcommand{\Erefs}[1]{Eqs.~(\ref{eq:#1})}

\newlength \standardfigwidth
\DeclareMathAlphabet{\matheub}{U}{eur}{m}{n}

\newcounter{exercise}
{\addtocounter{exercise}{1}\begin{center}\begin{minipage}{0.8\linewidth}\textbf{Exercise
\arabic{exercise}:}\begin{itshape}}
{\end{itshape}\end{minipage}\end{center}}

\makeatletter
\newcommand{\creat}[3][]{\@ifempty{#1}{#2^{\dagger}}{\left(#2^{\dagger}\right)^{#1}}\@ifempty{#3}{}{\!(#3)}}

\newcommand{\creatDoi}[3][]{\@ifempty{#1}{\tilde{#2}}{\left(\tilde{#2}\right)^{#1}}\@ifempty{#3}{}{(#3)}}

\newcommand{\annih}[3][]{#2\@ifempty{#1}{}{^{#1}}\@ifempty{#3}{}{(#3)}}

\makeatother

\newcommand{\hot}{\text{h.o.t.}}

% ...

\newlength{\bibmarkkeyAleft}

\newlength{\bibmarkkeyBleft}

\newlength{\bibmarkkeyCleft}

\newlength{\bibmarkkeyDleft}

\tikzset{
xxtsubstrate/.style={decorate, 
line width=1pt,
draw=olive, 
decoration=snake, 
segment amplitude=0.75mm, 
line after snake=0.25mm,
line before snake=0.25mm
},
tsubstrate/.style={decorate, 
line width=1pt,
draw=olive, 
decoration=snake, 
segment amplitude=0.5mm, 
segment length=5pt,
segment amplitude=0.2mm, 
line after snake=1mm,
line before snake=1mm
},
Bsubstrate/.style={decorate, 
line width=1pt,
draw=olive, 
decoration=snake,
segment length=5pt,
segment aspect=0,
segment amplitude=0.5mm, 
line after snake=0mm,
line before snake=0mm
},
substrate/.style={decorate, 
line width=1pt,
draw=olive, 
snake=snake, 
decoration=snake, 
segment length=5pt,
segment amplitude=0.5mm, 
line after snake=0.5mm,
line before snake=0.5mm
},
activity/.style={very thick,draw=red,postaction={decorate},
decoration={markings,mark=at position .5 with
{\arrow[draw=red]{>}}}},
cactivity/.style={very thick,draw=yellow,postaction={decorate},
decoration={markings,mark=at position .5 with
{\arrow[draw=yellow]{<}}}},
tactivity/.style={thick,draw=red,postaction={decorate},
decoration={markings,mark=at position .5 with
{\arrow[draw=red]{>}}}},
tEPSactivity/.style={thick,draw=red,postaction={decorate},
decoration={markings,mark=at position .55 with
{\arrow[draw=red]{>}}}},
tAactivity/.style={thick,draw=red},
Aactivity/.style={very thick,draw=black},
Bactivity/.style={very thick,draw=black,dashed},
dotactivity/.style={very thick,draw=black,dotted},
Cactivity/.style={very thick,draw=black,decorate,decoration=snake},
Baractivity/.style={very thick,draw=black},
tSactivity/.style={thick,draw=red,postaction={decorate},
decoration={markings,mark=at position .7 with
{\arrow[draw=red]{>}}}},
Sactivity/.style={very thick,draw=red,postaction={decorate},
decoration={markings,mark=at position .7 with
{\arrow[draw=red]{>}}}}
}
\tikzset{
  terminal/.style = {
    rectangle,
    align = center,
    minimum size = 6mm,
    rounded corners = 3mm,
    very thick,
    draw = black!50,
    top color = white,
    bottom color = black!20,
  }
}

\newcommand{\Abareprop}[2]{\tikz[baseline=-2.5pt]{
\draw[Aactivity] (0,0) -- +(-1.1,0) node[at start,above] {$#2$} node[at end,above]{$#1$} ;}}

\newcommand{\Bbareprop}[2]{\tikz[baseline=-2.5pt]{
\draw[Cactivity] (0,0) -- +(-1.1,0) node[at start,above] {$#2$} node[at end,above]{$#1$};}}

\newcommand{\Avertex}[4]{\tikz[baseline=-2.5pt]{
\draw[Aactivity] (0,0.3) -- +(-1,0) node[at start,above] {$#2$} node[at end,above]{$#1$} ;
\draw[Cactivity] (0,-0.3) -- +(-1,0) node[at start,below] {$#4$} node[at end,below]{$#3$};
\draw[Bactivity] (-0.5,0.3) -- +(0,-0.6);
}}

\newcommand{\twoDthreelegA}[3]{\tikz[baseline=-2.5pt]{
\draw[Aactivity] (0,0.3) -- +(-1,0) node[at start,above] {$#2$} node[at end,above]{$#1$} ;
\draw[Cactivity] (0,-0.3) -- +(-0.5,0) node[at start,below] {$#3$};
\draw[Bactivity] (-0.5,0.3) -- +(0,-0.6);
}}

\newcommand{\twoDthreelegB}[3]{\tikz[baseline=-2.5pt]{
\draw[Aactivity] (0,0.3) -- +(-0.5,0) node[at start,above] {$#2$};
\draw[Cactivity] (0,-0.3) -- +(-1,0) node[at start,below] {$#3$} node[at end,below]{$#1$};
\draw[Bactivity] (-0.5,0.3) -- +(0,-0.6);
}}

\newcommand{\noloop}{\tikz[baseline=-2.5pt]{
\draw[Aactivity] (0,0.3) -- +(-0.8,0) node[at end,above]{$x$} ;
\draw[Cactivity] (0,-0.3) -- +(-0.8,0) node[at end,below]{$y$};
}}

\newcommand{\zeroloop}{\tikz[baseline=-2.5pt]{
\draw[Aactivity] (0,0.3) -- +(-1,0) node[at end,above]{$x$} ;
\draw[Cactivity] (0,-0.3) -- +(-1,0) node[at end,below]{$y$};
\draw[Bactivity] (-0.3,0.3) -- +(0,-0.6);
}}

\newcommand{\oneloop}{\tikz[baseline=-2.5pt]{
\draw[Aactivity] (0,0.3) -- +(-1.5,0) node[at end,above]{$x$} ;
\draw[Cactivity] (0,-0.3) -- +(-1.5,0) node[at end,below]{$y$};
\draw[Bactivity] (-0.3,0.3) -- +(0,-0.6);
\draw[Bactivity] (-0.9,0.3) -- +(0,-0.6);
}}

\newcommand{\twoloop}{\tikz[baseline=-2.5pt]{
\draw[Aactivity] (0,0.3) -- +(-2.1,0) node[at end,above]{$x$} ;
\draw[Cactivity] (0,-0.3) -- +(-2.1,0) node[at end,below]{$y$};
\draw[Bactivity] (-0.3,0.3) -- +(0,-0.6);
\draw[Bactivity] (-0.9,0.3) -- +(0,-0.6);
\draw[Bactivity] (-1.5,0.3) -- +(0,-0.6);
}}

\tikzset{
    position label/.style={
       below = 3pt,
       text height = 1.5ex,
       text depth = 1ex
    },
   brace/.style={
     decoration={brace, mirror},
     decorate
   }
}

\newcommand{\jminusoneloop}{
\tikz[baseline=-2.5pt]{
\draw[Aactivity] (0,0.3) -- +(-1.6,0) node[at end,above]{$n$} ;
\draw[Cactivity] (0,-0.3) -- +(-1.6,0) node[at end,below]{$-n$};
\draw[Bactivity] (-0.3,0.3) -- +(0,-0.6);
\draw[Bactivity] (-0.9,0.3) -- +(0,-0.6);
\node at (0,0) [right] {\ldots};
\node at (0,0.3) [right] {\ldots};
\node at (0,-0.3) [right] {\ldots};
\draw[Aactivity] (1.8,0.3) -- +(-1.1,0) node[at start,above]{$_0$} ;
\draw[Cactivity] (1.8,-0.3) -- +(-1.1,0) node[at start,below]{$_0$};
\draw[Bactivity] (1.5,0.3) -- +(0,-0.6);
\draw[Bactivity] (0.9,0.3) -- +(0,-0.6);
  \draw [brace] (1.6,0.8) -- (-1,0.8) node [position label, pos=0.5, above] {$j$};
}}

\newcommand{\jloop}{
\tikz[baseline=-2.5pt]{
\draw[Aactivity] (0,0.3) -- +(-3.5,0) node[at end,above]{$n,\omega$} ;
\draw[Cactivity] (0,-0.3) -- +(-3.5,0) node[at end,below]{$-n,-\omega$};
\draw[Bactivity] (-0.3,0.3) -- +(0,-0.6);
\draw[Bactivity] (-0.9,0.3) -- +(0,-0.6);
\draw[Bactivity] (-2.1,0.3) -- +(0,-0.6);
\node at (-2.1,0.55) [left]{$-n$};
\node at (-2.1,0.55) [right]{$m$};
\node at (-2.1,-0.55) [left]{$n$};
\node at (-2.2,-0.55) [right]{$-m$};
\node at (0,0) [right] {\ldots};
\node at (0,0.3) [right] {\ldots};
\node at (0,-0.3) [right] {\ldots};
\draw[Aactivity] (1.8,0.3) -- +(-1.1,0) node[at start,above]{$_0$} ;
\draw[Cactivity] (1.8,-0.3) -- +(-1.1,0) node[at start,below]{$_0$};
\draw[Bactivity] (1.5,0.3) -- +(0,-0.6);
\draw[Bactivity] (0.9,0.3) -- +(0,-0.6);
  \draw [brace] (1.6,0.8) -- (-2.2,0.8) node [position label, pos=0.5, above] {$j+1$};
}}

\newcommand{\threeToTwoPointB}{\tikz[baseline=6pt]{
\draw[Aactivity] (0,0.3) -- +(-1.9,0) node[at end,above]{$x$} ;
\draw[Cactivity] (0,-0.3) -- +(-1.9,0) node[at end,below]{$y$};
\draw[Cactivity] (0,0.9) -- +(-1.3,0);
\draw[Bactivity] (-0.6,0.3) -- +(0,0.6);
\draw[Bactivity] (-1.3,0.3) -- +(0,0.6);
\draw[Bactivity] (-0.3,0.3) -- +(0,-0.6);
}}

\newcommand{\threeToTwoPointA}{\tikz[baseline=6pt]{
\draw[Aactivity] (0,0.3) -- +(-1.5,0) node[at end,above]{$x$} ;
\draw[Cactivity] (0,-0.3) -- +(-1.5,0) node[at end,below]{$y$};
\draw[Cactivity] (0,0.9) -- +(-1,0);
\draw[Bactivity] (-0.3,0.3) -- +(0,0.6);
\draw[Bactivity] (-0.6,0.3) -- +(0,-0.6);
\draw[Bactivity] (-1,0.3) -- +(0,0.6);
}}

\newcommand{\threeToTwoPointC}{\tikz[baseline=6pt]{
\draw[Aactivity] (0,0.3) -- +(-1.9,0) node[at end,above]{$x$} ;
\draw[Cactivity] (0,-0.3) -- +(-1.9,0) node[at end,below]{$y$};
\draw[Cactivity] (0,0.9) -- +(-1,0);
\draw[Bactivity] (-0.3,0.3) -- +(0,0.6);
\draw[Bactivity] (-1,0.3) -- +(0,0.6);
\draw[Bactivity] (-1.4,0.3) -- +(0,-0.6);
}}

\newcommand{\threeToTwoPointD}{\tikz[baseline=6pt]{
\draw[Aactivity] (0,0.3) -- +(-1.3,0) node[at end,above]{$x$} ;
\draw[Cactivity] (0,-0.3) -- +(-1.3,0) node[at end,below]{$y$};
\draw[Cactivity] (0,0.9) -- +(-0.6,0);
\draw[Bactivity] (-0.3,0.3) -- +(0,-0.6);
\draw[Bactivity] (-0.6,0.3) -- +(0,0.6);
}}

\newcommand{\threeToTwoPointE}{\tikz[baseline=6pt]{
\draw[Aactivity] (0,0.3) -- +(-1.3,0) node[at end,above]{$x$} ;
\draw[Cactivity] (0,-0.3) -- +(-1.3,0) node[at end,below]{$y$};
\draw[Cactivity] (0,0.9) -- +(-0.3,0);
\draw[Bactivity] (-0.3,0.3) -- +(0,0.6);
\draw[Bactivity] (-0.6,0.3) -- +(0,-0.6);
}}

\newcommand{\threebareprop}{\tikz[baseline=6pt,scale=0.9]{
\draw[Aactivity] (0,0.3) -- +(-1,0) node[at end,above]{$x$} ;
\draw[Cactivity] (0,-0.3) -- +(-1,0) node[at end,below]{$y$};
\draw[Cactivity] (0,0.9) -- +(-1,0) node[at end,above]{$y'$};
}}

\newcommand{\threepropA}{\tikz[baseline=6pt,scale=0.9]{
\draw[Aactivity] (0,0.3) -- +(-1,0) node[at end,above]{$x$} ;
\draw[Cactivity] (0,-0.3) -- +(-1,0) node[at end,below]{$y$};
\draw[Cactivity] (0,0.9) -- +(-1,0) node[at end,above]{$y'$};
\draw[Bactivity] (-0.3,0.3) -- +(0,-0.6);
}}

\newcommand{\threepropB}{\tikz[baseline=6pt,scale=0.9]{
\draw[Aactivity] (0,0.3) -- +(-1,0) node[at end,above]{$x$} ;
\draw[Cactivity] (0,-0.3) -- +(-1,0) node[at end,below]{$y$};
\draw[Cactivity] (0,0.9) -- +(-1,0) node[at end,above]{$y'$};
\draw[Bactivity] (-0.3,0.3) -- +(0,0.6);
}}

\newcommand{\threepropAA}{\tikz[baseline=6pt,scale=0.9]{
\draw[Aactivity] (0,0.3) -- +(-1.5,0) node[at end,above]{$x$} ;
\draw[Cactivity] (0,-0.3) -- +(-1.5,0) node[at end,below]{$y$};
\draw[Cactivity] (0,0.9) -- +(-1.5,0) node[at end,above]{$y'$};
\draw[Bactivity] (-0.3,0.3) -- +(0,-0.6);
\draw[Bactivity] (-0.9,0.3) -- +(0,-0.6);
}}

\newcommand{\threepropBB}{\tikz[baseline=6pt,scale=0.9]{
\draw[Aactivity] (0,0.3) -- +(-1.5,0) node[at end,above]{$x$} ;
\draw[Cactivity] (0,-0.3) -- +(-1.5,0) node[at end,below]{$y$};
\draw[Cactivity] (0,0.9) -- +(-1.5,0) node[at end,above]{$y'$};
\draw[Bactivity] (-0.3,0.3) -- +(0,0.6);
\draw[Bactivity] (-0.9,0.3) -- +(0,0.6);
}}

\newcommand{\threepropAB}{\tikz[baseline=6pt,scale=0.9]{
\draw[Aactivity] (0,0.3) -- +(-1.3,0) node[at end,above]{$x$} ;
\draw[Cactivity] (0,-0.3) -- +(-1.3,0) node[at end,below]{$y$};
\draw[Cactivity] (0,0.9) -- +(-1.3,0) node[at end,above]{$y'$};
\draw[Bactivity] (-0.3,0.3) -- +(0,-0.6);
\draw[Bactivity] (-0.6,0.3) -- +(0,0.6);
}}

\newcommand{\threepropBA}{\tikz[baseline=6pt,scale=0.9]{
\draw[Aactivity] (0,0.3) -- +(-1.3,0) node[at end,above]{$x$} ;
\draw[Cactivity] (0,-0.3) -- +(-1.3,0) node[at end,below]{$y$};
\draw[Cactivity] (0,0.9) -- +(-1.3,0) node[at end,above]{$y'$};
\draw[Bactivity] (-0.3,0.3) -- +(0,0.6);
\draw[Bactivity] (-0.6,0.3) -- +(0,-0.6);
}}

\usetikzlibrary{decorations.pathmorphing}
\usetikzlibrary{decorations.markings}

 \newcommand{\AD}{\mathcal{D}}
 
 \newcommand{\AH}{\mathcal{H}}

 \newcommand{\action}{\mathcal{A}}
 \newcommand{\actionPert}{\action_{\text{\tiny pert}}}
 
  \usetikzlibrary{snakes}
  \newcommand{\ep}{\dot{S}}
\newcommand{\eqhat}{\mathrel{\hat{=}}}
\usepackage{CJK}
\newcommand{\yset}{\yvec}

\newcommand{\Nsum}{\sum_{i=1}^N}
\newcommand{\tildepsii}{\tildepsi}
\newcommand{\psii}{\psi}
\newcommand{\tildephi}{\widetilde\phi}
\newcommand{\tildepsi}{\widetilde{\psi}}

\newcommand{\zilname}{\begin{CJK*}{UTF8}{gbsn}(张子洛)\end{CJK*}}
\allowdisplaybreaks

\newcommand{\titleText}{Entropy production of non-reciprocal interactions}
\title{\titleText}

\CJKnospace
\begin{document}

\title{Entropy production of non-reciprocal interactions}

\author{Ziluo Zhang \zilname }
\affiliation{Department of Mathematics, Imperial College London, London SW7 2AZ, UK}
\author{Rosalba Garcia-Millan}
\email{rg646@cam.ac.uk}
\affiliation{DAMTP, Centre for Mathematical Sciences, University of Cambridge, Cambridge CB3 0WA, UK}
\affiliation{St John's College, University of Cambridge, Cambridge CB2 1TP, UK}
\date{\today}

\begin{abstract}
Non-reciprocal interactions are present in many systems out of equilibrium.
The rate of entropy production is a measure that quantifies the time irreversibility of a system, and
thus how far it is from equilibrium.
In this work, we introduce a non-motile active particle system where activity originates from
asymmetric, pairwise interaction forces that result in an injection of energy at the microscopic scale.
We calculate stationary correlation functions and entropy production rate in three
exactly solvable cases,
and analyse a more general case in a perturbation theory as an expansion in weak interactions
using a fully microscopic description.
Our results show that equilibrium may be recovered by adjusting the diffusion constants despite
non-reciprocity, revealing an equivalence in the absolute amplitude of the force and diffusivity.
We support our analytical results with numerical simulations.
\end{abstract}

\maketitle

\begin{figure}
    \centering
    \includegraphics[width=0.49\textwidth]{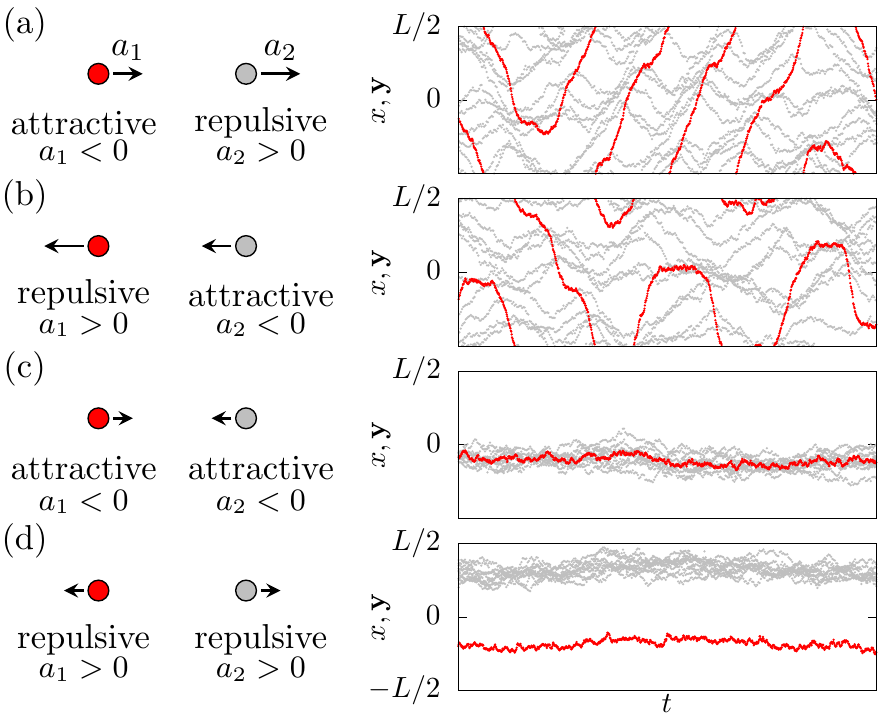}
   \caption{{\bf Particle trajectories $(x,\yvec)$} described by \Eref{L_eqs} with 
   one particle of species A (red) and
   $N=10$ particles of species B (grey),
    sinusoidal interacting potentials,
    equal diffusivities and zero drifts.
    The sign of potential amplitudes $a_1$ and $a_2$ are indicated on the left.
    Interactions are non-reciprocal in (a) and (b) and reciprocal in (c) and (d).
    In this example, since $D_1=D_2$,  
    non-reciprocal systems are active
    and
    reciprocal systems are passive.
   }
    \label{fig:traj03}
\end{figure}

\textit{Introduction.}---Non-reciprocal interactions are those that do not obey Newton's
third law (\textit{actio} equal \textit{reactio}).
These 
 generate intrinsically out-of-equilibrium dynamics 
\cite{Loos_2020,GodrecheLuck:2019,KanoETAL:2017}
and are often invoked to 
model striking dynamical pattern formation such as flocking \cite{Cavagna_2018,NagyETAL:2010},
worming \cite{Deblais:2020} or travelling states \cite{YouETAL:2020}.
Common mechanisms that break reciprocity are
(anti-)alignment  \cite{FruchartETAL:2021} and vision cone interactions
\cite{Durve:2018,Barberis:2016,LoosKlappMartynec:2022}.
Non-reciprocal interactions between two species play a crucial role in 
numerous biological processes, such
as predator-prey dynamics \cite{Redner_1999,MeredithETAL:2020}
or collective cell migration \cite{Stramer:2017}.
Non-reciprocity is found in attractive-repulsive interactions giving rise,
for example, to the chase-and-run dynamics present between a dog and
a herd of sheep \cite{Stramer:2017}:
while the dog runs toward the sheep,
sheep fearfully run away from the dog,
following a trajectory such as in Fig.~\ref{fig:traj03}(a).

Systems with non-reciprocal interactions are generally out of equilibrium,
which is characterised by the breakdown of time-reversal symmetry.
The rate of entropy production has become, over the last decade,
a prominent measure of irreversibility that quantifies 
how far from equilibrium a system is
\cite{PruessnerGarcia-Millan:2022,
Gaspard:2007,Cocconi:2020,AlstonCocconiBertrand:2022}.
Measuring and calculating the entropy production rate is, therefore, key to 
understanding the non-equilibrium behaviour of a system
and poses a major theoretical challenge in the case of
interacting many-particle active systems.
Recent studies have focused on the thermodynamic properties of non-reciprocity
in a system described by coupled, linear Langevin equations 
\cite{Loos_2020,GodrecheLuck:2019,BrandenbourgerETAL:2019}.
Reference \cite{Loos_2020} showed that there exist systems with non-reciprocal interactions
that do satisfy detailed balance.
However, a fully microscopic calculation of the entropy production of a
many-particle system with non-linear and non-reciprocal interaction 
forces is lacking. 
The power of microscopic theories in the study of active particle systems lies in
the direct link they establish between agents' properties and their 
emergent phenomena.
Deriving exact results that serve as benchmark is thus of
paramount importance \cite{Cocconi:2020,PruessnerGarcia-Millan:2022}. 

In this letter, we introduce a non-motile
active particle system that is driven
out of equilibrium by
non-reciprocal interactions between one particle and the rest
\cite{AlstonETAL:2022}.
We use a fully microscopic description to calculate correlation 
functions and entropy production.
We consider three exactly solvable cases and a fourth, more general case using
a perturbative field theory.
Our perturbative approach provides a  systematic framework
to calculate correlation functions and the short-time propagator, 
both of which are essential to calculate the entropy production rate
\cite{ZhangPruessner:2021,Garcia-MillanPruessner:2021, PruessnerGarcia-Millan:2022,BotheCocconiZhenPruessner:2022}.
Our analytical results are compared with the average rate of dissipated heat
along the stochastic trajectories of the particles in Sekimoto's framework \cite{Loos_2020, 
Sekimoto:2010,GhosalBisker:2022,RoldanETAL:2021}, showing good 
agreement
within the domain where our perturbative treatment holds.
We address the regime of small to intermediate
particle numbers, and establish the path from microscopic dynamics
to emergent effective interactions.

Following the example of chase-and-run dynamics described above,
we consider a one-dimensional system with two types of particles: 
one particle of species A at position $x$
and $N$ particles of species $B$ at $\yvec=(y_1,\ldots,y_N)$,
with diffusion constants $D_1$ and $D_2$, 
and drifts $u_1 $ and $u_2$, that live on a
ring of length $L$. 
The interaction forces
are mediated by the periodic and bounded pair potentials $V_1$ and $V_2$, resulting in the
additional effective drifts 
$-\sum_i\partial_x V_1(x-y_i)$ for particle A
and $-\partial_{y_i} V_2(y_i-x)$ for each particle B.
According to 
the structure of these pair interactions, B particles do not directly interact with
each other, although their trajectories are effectively coupled by means of particle A,
as we discuss below.

The system dynamics are described by the coupled, overdamped Langevin equations,
\begin{subequations}
\elabel{L_eqs}
\begin{align}
    \dot{x}&=u_1-\Nsum V'_1(x-y_i)+\xi(t) \ ,\\
    \dot{y}_i&=u_2-V'_2(y_i-x)+\xi_i(t) \ ,
\end{align}
\end{subequations}
where $\xi$ and $\xi_i$ are Gaussian white noises, with correlators 
$\ave{\xi(t) \xi(t')}=2D_1\delta(t-t')$
and
$\ave{\xi_i(t) \xi_j(t')}=2D_2\delta_{i,j}\delta(t-t')$.
We assume Boltzmann constant and mobility 
to be unity,
and therefore deem
the diffusion constants to essentially 
be temperatures.
The stationary probability currents of $x$ and $y_i$
are
$        J_x =  - \left(-u_1+ \Nsum V_1'(x-y_i) + D_1\partial_x
        \right) P(x,\yset) $ 
        and
$        J_{y_i} =  - \left(-u_2+ V_2'(y_i-x) + D_2\partial_{y_i}
        \right) P(x,\yset) 
$,
where $P(x,\yvec)$ is the joint probability density of state $x, \yvec$ \cite{Risken:1989}.
If interactions are reciprocal, namely the interaction potentials satisfy 
$V_1(\ell)=V_2(\ell)\equiv V(\ell) $ with $\ell=x-y$, then forces are conservative because
 \Eref{L_eqs} can be derived from
the Hamiltonian 
$\AH = -u_1x+ \Nsum\left(
    -u_2y_i + V(x-y_i)
    \right)$.
    Conversely, non-reciprocal interactions result in non-conservative forces.
Fig.~\ref{fig:traj03} shows particle trajectories described by
\Eref{L_eqs} with different combinations of attractive and repulsive
potentials, illustrating the breakdown of time-reversal symmetry 
in the active systems in Figs.~\ref{fig:traj03}(a) and \ref{fig:traj03}(b),
in contrast to the passive systems in Figs.~\ref{fig:traj03}(c) and \ref{fig:traj03}(d).
We simulate particle trajectories described by \Eref{L_eqs} using standard Brownian
dynamics simulations.

We calculate
the entropy production to quantify the irreversibility of this system
\cite{Gaspard:2007,Cocconi:2020,AlstonCocconiBertrand:2022}.
In Gaspard's framework \cite{Gaspard:2007}, the entropy production
of a Markov process is
\begin{align}
\elabel{def_entropy_production}
    \ep(t)= & \lim_{\tau\rightarrow 0}
    \frac{1}{(N!)^2}
    \int_0^L \dint{x} \dint{x'}  \Ndint{y}  \Ndint{y'} 
    \nonumber \\ &
    \Bigg\{
    P(x,\yvec;t) \dot{W}(x,\yvec \rightarrow x',\yvec';\tau)
    \nonumber \\ &
    \times \ln \left(
    \frac
    {P(x,\yvec;t) W(x,\yvec \rightarrow x',\yvec';\tau)}
    {P(x',\yvec';t) W(x',\yvec' \rightarrow x,\yvec;\tau)}
    \right) \Bigg\} \ ,
\end{align}
where 
$W(x,\yvec \rightarrow x',\yvec';\tau)$ is the transition probability for the system to go from state $x,\yset$ to state
$x',\yset'$ in an interval of time $\tau$.
The Gibbs factor $1/N!$ in each integral over  
$\yvec$ in \Eref{def_entropy_production} accounts for the phase space of the positions of B particles being
that of
indistinguishable particles in the continuum, where the probability that two particles are found at the same position has  zero measure.

Using the framework established in \cite{PruessnerGarcia-Millan:2022}, 
the entropy production in \Eref{def_entropy_production} in the stationary
state
 simplifies to
\begin{widetext}
\begin{align}
     \ep=&
     \int_0^L\dint{x} \dint{y}\bigg\{-V''_1(x-y)+\frac{1}{D_1}\left(\frac{u_1}{N}-V'_1(x-y)\right)^2
     -V''_2(y-x)+\frac{1}{D_2}\left(u_2-V'_2(y-x)\right)^2\bigg\}P^{(1+N)}_2(x,y)
     \nonumber\\
     &+\int_0^L\dint{x} \dint{y}\dint{y'}\frac{1}{D_1}\left(\frac{u_1}{N}-V'_1(x-y)\right)\left(\frac{u_1}{N}-V'_1(x-y')\right) P^{(1+N)}_3(x,y,y')
     \ ,
     \elabel{ep_many_sheep_simplified}
    \end{align}
\end{widetext}
which is derived in the Supplemental Material (SM). 
\Eref{ep_many_sheep_simplified} shows that the entropy production of this system depends
on the two-point $P^{(1+N)}_2$ and three-point $P^{(1+N)}_3$ correlation functions
\cite{PruessnerGarcia-Millan:2022}.
In the absence of an exact solution,
we calculate these correlation functions perturbatively,
using a Doi-Peliti field theory that captures the
microscopic dynamics of \Eref{L_eqs}.

\begin{figure}
	\centering
		\includegraphics[width=0.475\textwidth]{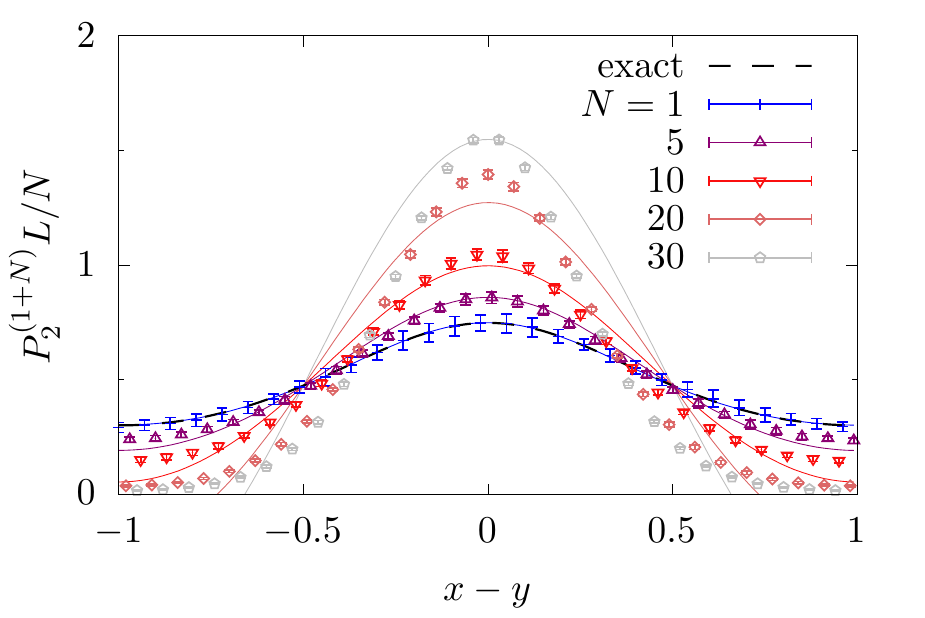}
	\caption{{\bf Rescaled two-point correlation function  $P^{(1+N)}_2(x,y)$ }
	of A and B particles at $x$ and $y$, with sinusoidal interaction potentials
	and $a_1=-0.2$, $a_2=-0.3$, $u_1=u_2=0$, $D_1=1$, $D_2=0.1$, $L=2$.
	Symbols show numerical estimates;
	dashed lines, exact result \Eref{two_point_corr_no_drift}; and
	solid lines, perturbative prediction (SM). 
	}
	\label{fig:two_point_correlation}
\end{figure}

\begin{figure}[t]
    \centering
    \includegraphics[width=0.5\textwidth]{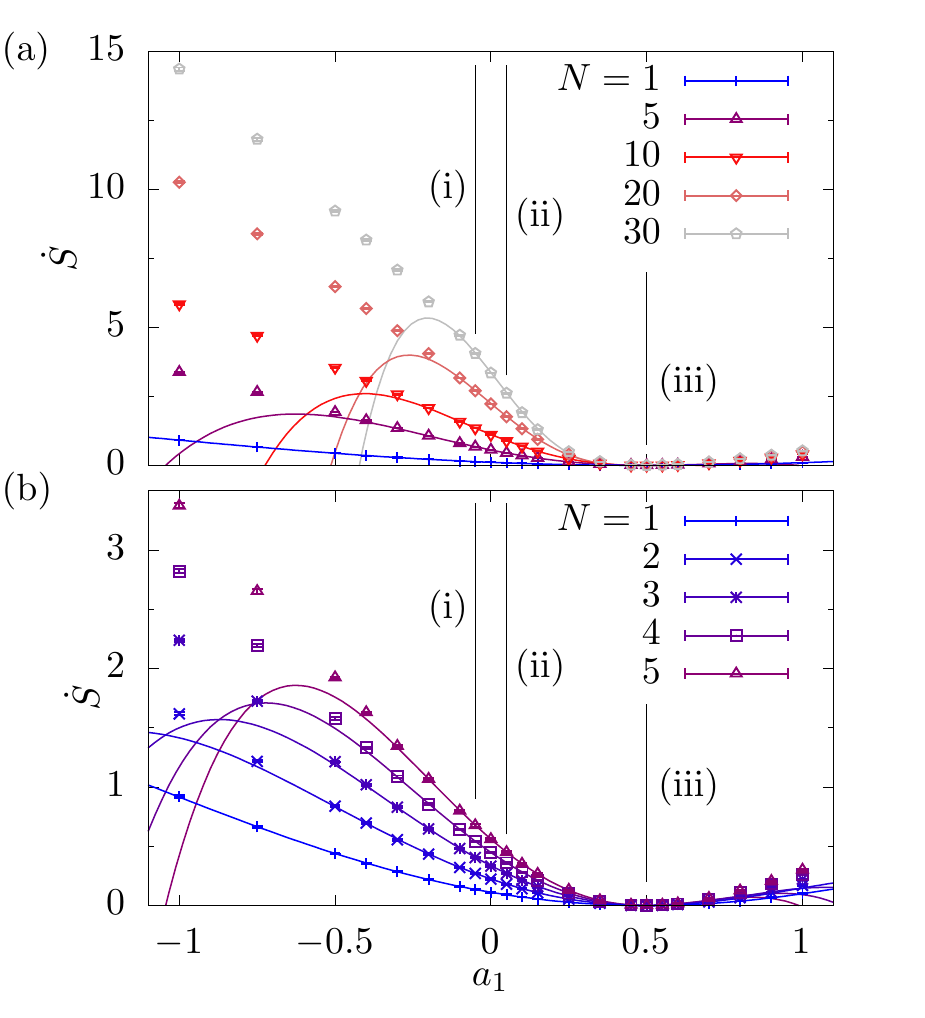}
\caption{     \label{fig:ep_chasing}
{\bf Entropy production $\ep$} varying $a_1$, 
(a) $N\in\{1,\ldots,30\}$ and (b) $N\in\{1,\ldots,5\}$,
and fixed $a_2=0.05$, $D_1=1$, $D_2=0.1$, $u_1=u_2=0$, $L=2$.
Symbols indicate numerical estimates using Sekimoto's framework, and
lines perturbative prediction in (SM). 
The cases (i) chase-and-run dynamics $a_1=-a_2$, (ii) reciprocal interactions $a_1=a_2$, and
(iii) equilibrium $a_1=a_2 D_1/D_2$, are indicated.
The system with reciprocal interactions is out of equilibrium because 
$D_1\neq D_2$, \Eref{detailed_balance_two}. 
}
\end{figure}

\textit{Two-particle system.}---The system with one particle of each species is tractable exactly.
Without self-propulsion, $u_1=u_2=0$, the two-point correlation function  
is the barometric formula 
\cite{Risken:1989,FehertoiPolackova:ToBePublished}
\begin{equation}
    P_2^{(2)}(x,y)=
    \frac{1}{L \norm}
    \exp{-\frac{V_1(x-y)+V_2(y-x)}{D_1+D_2}} \ ,
    \elabel{two_point_corr_no_drift}
\end{equation}
with  $\norm$ such that 
$
\int_{0}^L\dint{x}\dint{y}P^{(2)}_{2}(x,y)=1
$, which is represented with dashed lines in Fig.~\ref{fig:two_point_correlation}.
The three-point correlation function is
naturally $P_3^{(2)}=0$.
Setting the probability currents  to zero, $J_x=0$ and $J_y=0$,
gives the condition for detailed balance for this system,
\begin{equation}
  \frac{V'_1(x-y)}{D_1}=- \frac{V'_2(y-x)}{D_2} \ ,
  \elabel{detailed_balance_two}
\end{equation}
showing an equivalent role in the 
amplitude of interaction force and diffusion.
\emph{The system
can therefore be out of equilibrium due to particle interactions
even in the absence of self-propulsion}.
The detailed balance condition in \Eref{detailed_balance_two}
is consistent with the detailed balance condition found in
\cite{Loos_2020}
for the special case of harmonic interaction potentials.
Using \Eref{two_point_corr_no_drift} in \Eref{ep_many_sheep_simplified},
the stationary entropy production is
\begin{equation}
    \ep=\int_{0}^L\dint\ell \frac{\rho(\ell)}{D_1+D_2}
    \left(\sqrt{\frac{D_2}{D_1}}V'_1(\ell)+\sqrt{\frac{D_1}{D_2}}V'_2(-\ell)\right)^2 \ ,
    \elabel{EP_zero_drift}
\end{equation}
where $\rho(x-y)= LP_2^{(2)}(x,y)$.
The entropy production is zero only if \Eref{detailed_balance_two} is satisfied.
Fig.~\ref{fig:ep_chasing} shows $\ep$ for the two-particle case, $N=1$,
for sinusoidal interaction potentials.

\textit{Equilibrium system.}---Imposing the detailed balance 
condition in \Eref{detailed_balance_two} and zero
drifts $u_1=u_2=0$,
the stationary joint probability density is
\begin{equation}
    P(x,\yvec)= 
        \frac{1}{L \norm}
    \exp{-\Nsum\frac{V_1(x-y_i)+V_2(y_i-x)}{D_1+D_2}}
    \elabel{DB_correlation} \ ,
\end{equation}
with normalisation $\norm$ such that 
$
\int_{0}^L\dint{x}\Ndint{y}P(x,\yvec)=N!
$.
The probability currents 
vanish, implying that the detailed balance condition in \Eref{detailed_balance_two}
generalises to the many-particle system.

\begin{figure*}
	\centering
		\includegraphics[width=0.9\textwidth]{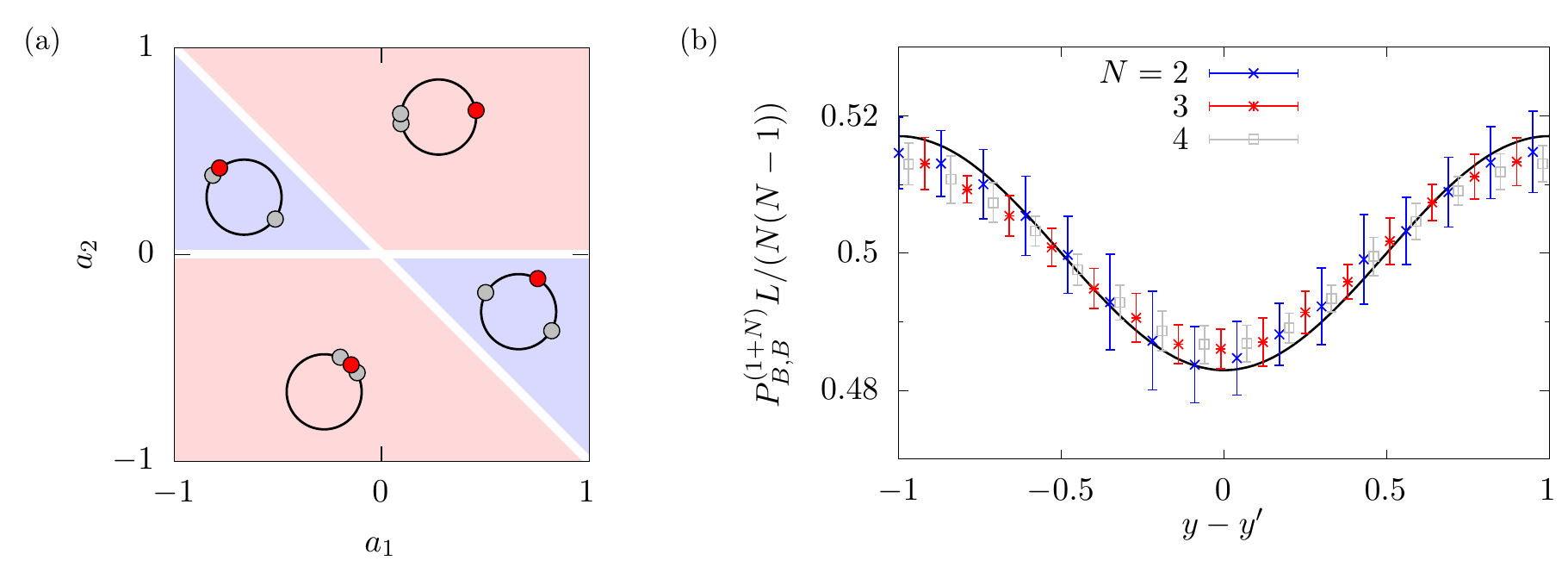}	
	\caption{
	{\bf Effective interactions} between particles of species B.
	(a) Phase diagram showing effective attraction (red) and repulsion (blue)
	varying $a_1$ and $a_2$, as predicted by the amplitude 
	of the two-point correlation function  $P^{(1+N)}_{B,B}(y,y')$
	of two B particles at
	$y$ and $y'$, in the presence of
	an A particle,
	insets show frequent configurations in each regime. 
	(b) Rescaled  $P^{(1+N)}_{B,B}$, with sinusoidal interaction potentials, and
	$a_1=-0.2$, $a_2=0.15$, $D_1=1$, $D_2=0.1$, $u_1=u_2=0$, $L=2$.
	Symbols indicate numerical estimates, and solid line,
	perturbative result (SM). 
	}
	\label{fig:sheep_sheep_corr}
\end{figure*}

\textit{Chase and run.}---The limiting case where interaction forces 
between species are attractive-repulsive with equal magnitude and equal directions
is of particular interest both for its biological motivation \cite{Stramer:2017},
as well as for its mathematical tractability.
Assuming that interaction potentials satisfy $V_1(\ell)=-V_2(-\ell)\equiv V(\ell)$,
the two-particle system is uniformly distributed, \Eref{two_point_corr_no_drift},
because the distance $\ell=x-y$ diffuses in the comoving frame, and
it is the centre of mass that spins with velocity $-V'(\ell)$.
By the same reasoning, the stationary joint probability density stays uniform
when adding new particles to the system.
By marginalisation,
the $n$-point correlation 
function is
\begin{equation}
\elabel{n_point_uniform}
    P_n^{(1+N)}(x,y_1,\ldots,y_{n-1})=\frac{N!}{(N-n+1)!L^{n}} \ , 
\end{equation}
for $n\geq 1$.
The entropy production in \Eref{ep_many_sheep_simplified} is, thus,
\begin{equation}
    \ep=\frac{N}{L } \left(\frac{1}{D_1}+\frac{1}{D_2} \right)\int_0^L\dint{\ell}
    \left({V'(\ell)}\right)^2 \ ,
    \elabel{EPchaserun}
\end{equation}
yielding an exact result for arbitrary particle number $N$.
\Eref{EPchaserun} 
shows that the system with attractive-repulsive interactions
is out of equilibrium independently of the magnitude of interaction forces,
as indicated in Fig.~\ref{fig:ep_chasing}.
Indeed, the \emph{chaser}
tends to run behind the \emph{chased}, exhibiting the 
breakdown of time-reversal symmetry.

\textit{Sinusoidal interaction potentials.}---To explore the parameter space of
attractive and repulsive interactions, 
we consider sinusoidal potentials 
$V_1(\ell)=a_1\cos{k_1 \ell}$ and 
    $V_2(\ell)=a_2\cos{k_1 \ell}$,
with $k_1=2\pi/L$.
The signs of $a_1$ and $a_2$ determine whether interactions are 
attractive or repulsive: negative amplitudes result in attraction, whereas
positive amplitudes result in repulsion, as shown in Fig.~\ref{fig:traj03}.

We derive the two-point and three-point correlation functions,
in a perturbative expansion valid at small amplitudes $a_1$ and $a_2$
compared to the diffusion constants $D_1$ and $D_2$ in the SM.
In Fig.~\ref{fig:two_point_correlation}, we show the two-point correlation function 
$P_2^{(1+N)}(x,y)$ of the A particle at position $x$ and one out of $N$ particles
of species B at $y$,
as a function of the distance $x-y$, varying $N$.
Due to attractive interactions, short distances are favoured over longer distances,
which is more prominent as $N$ increases.
Fig.~\ref{fig:two_point_correlation} shows good agreement for small
to intermediate $N$ and
illustrates how the deviation of the 
numerical estimates from the perturbative prediction increases with $N$.

Using the correlation functions $P_2^{(1+N)}$ and $P_3^{(1+N)}$
in \Eref{ep_many_sheep_simplified} we obtain the
entropy production $\ep$, derived in the SM, which is shown in
Fig.~\ref{fig:ep_chasing} as a function of $a_1$ (solid lines).
The behaviour of $\ep$ is shown
for two ranges of $N$:  the small to large particle number in Fig.~\ref{fig:ep_chasing}(a),
and the small to intermediate
particle numbers  in Fig.~\ref{fig:ep_chasing}(b).
Following Sekimoto's
framework \cite{Loos_2020, Sekimoto:2010,GhosalBisker:2022,RoldanETAL:2021},
in Fig.~\ref{fig:ep_chasing} we compare
the perturbative result for $\ep$
with the total entropy production
$\ep = {\dot{Q}_x}/{\TC_1} + \Nsum {\dot{Q}_{y_i}}/{\TC_2}$
estimated numerically from the average rate of dissipated energy
along particle trajectories,
$\dot{Q}_x = 
        \ave{\left(
        \dot{x}(t) - \xi(t)
        \right)
        \circ \plaind x(t)}/{\plaind t}$
and
$ \dot{Q}_{y_i} = 
        \ave{\left(
        \dot{y_i}(t) - \xi_i(t)
        \right)
        \circ \plaind {y_i}(t)}/{\plaind t}$, 
where $\circ$ denotes the Stratonovich
product and
where the temperatures are $\TC_1=D_1/(k_{\text{B}}\mu)$,
and $\TC_2=D_2/(k_{\text{B}}\mu)$,
with Boltzmann constant $k_{\text{B}}$ and mobility $\mu$
set to unity.

Fig.~\ref{fig:ep_chasing} shows that the entropy production $\ep$ increases smoothly for
amplitudes away from the equilibrium $a_1=a_2 D_1/D_2$, and for
increasing $N$.
It further shows that $\ep$ grows faster in the attractive-repulsive regime 
($a_1<0$) than in the repulsive-repulsive regime ($a_1>0$).
The cases of chase-and-run dynamics, reciprocal interactions and equilibrium
are indicated. Here, since the diffusion constants are different, $D_1\neq D_2$,
the system with reciprocal interactions does not satisfy the condition for
detailed balance in \Eref{detailed_balance_two} and is thus out of equilibrium.
The agreement is excellent between numerical results and
perturbative predictions within
the validity regime, that is for $|a_1|$ and $N$ sufficiently small. 
Increasing
either of those parameters
leads to larger deviation of the numerical estimates
from the perturbative prediction. 

\textit{Effective interactions.}---Although particles of species B do not directly interact
with each other according to 
\Eref{L_eqs}, their trajectories are correlated
through the particle of species A, 
as illustrated in Fig.~\ref{fig:traj03}. In Figs.~\ref{fig:traj03}(c) and (d),
for instance, B particles visibly display effective attraction.
Marginalising the three-point correlation function,
$\int_0^L\dint{x}P^{(1+N)}_3(x,y,y')=P_{B,B}^{(1+N)}(y,y')$,
reveals correlations between any
two particles of species B at positions $y$ and $y'$, 
that we identify as effective interactions, with a sign (attractive or repulsive) that depends 
on the amplitudes $a_1$ and $a_2$, derived in the SM.
We found that B particles display effective repulsion if
$a_1>-a_2>0$ or $-a_1>a_2>0$ 
(blue area in the phase diagram in Fig.~\ref{fig:sheep_sheep_corr}(a)),
no correlation if $a_1=-a_2$ or $a_2=0$ (white lines),
and effective attraction otherwise
(red area).
Based on this prediction, which is of second order in the perturbation
expansion, we explored the two-point correlation function $P_{B,B}^{(1+N)}$
for an example of effective repulsion in the regime $-a_1>a_2>0$,
see Fig.~\ref{fig:sheep_sheep_corr}(b),
which shows that long distances are more frequent than short distances
between any two B particles.
The underlying microscopic mechanism is the following:
the attraction felt by particle A is stronger than the repulsion felt by B particles,
so that A captures one of the B particles, while the rest escape.

\textit{Conclusions.}---
We have studied the role of non-reciprocal interactions in the environmental
entropy change using a fully microscopic description.
We have shown that non-reciprocal forces generally break detailed balance,
resulting in a local injection of energy and turning a system of
non-motile particles into an active one.
To our knowledge, this is the first microscopic field theory of an interacting, 
many-particle active system.
Our framework can be extended to estimate the extractable work in other 
physical systems, such as asymmetrically shaped objects immersed in 
active baths that have been proposed in the context of autonomous engines.

We are grateful to G.~Pruessner for introducing us to field theory and for seminal
discussions about the barometric formula.
We also thank S.~Loos, C.~Scalliet, M.~Cates,
T.~Agranov, P.~Pietzonka,
G.~Biroli, B.~Walter, L.~Cocconi and  Z.~Zhen for useful discussions.
RG-M was supported in part by the European Research Council under the EU's Horizon 2020 Programme (Grant number 740269), and acknowledges support from a St John's College Research Fellowship, University of Cambridge.

\bibliography{LSA}
\bibliographystyle{unsrt}

\onecolumngrid
\newpage
\newpage
\renewcommand{\theequation}{S-\arabic{equation}}
\setcounter{equation}{0}
\setcounter{page}{1}
\renewcommand{\thepage}{S-\arabic{page}}

\begin{center}
\textbf{\large Supplemental Material: \titleText}

{Ziluo Zhang and Rosalba Garcia-Millan}
\end{center}

\section{Entropy production}
\label{entropy_production}
We calculate the internal entropy production rate in \Eref{def_entropy_production}
of this Markov process via \cite{Gaspard:2007,Cocconi:2020,AlstonCocconiBertrand:2022,PruessnerGarcia-Millan:2022}.
At stationarity, the internal entropy production simplifies as the probability $P$ disappears from the logarithm \cite{Cocconi:2020, Garcia-MillanPruessner:2021, PruessnerGarcia-Millan:2022},
\begin{equation}
   \ep=\lim_{\tau\rightarrow 0}
    \frac{1}{(N!)^2}
    \int_0^L \dint{x} \dint{x'}  \Ndint{y}  \Ndint{y'} 
   P(x,\yvec) \dot{W}(x,\yvec \rightarrow x',\yvec';\tau)
    \ln \left(
    \frac
    { W(x,\yvec \rightarrow x',\yvec';\tau)}
    { W(x',\yvec' \rightarrow x,\yvec;\tau)}
    \right)\elabel{stationary_entropy_production}  \ ,
\end{equation}
where $P(x,\yvec)=\lim_{t\rightarrow \infty}P(x,\yvec;t)$ is the stationary density.
Following the approach in \cite{PruessnerGarcia-Millan:2022}
we obtain the transition rate $\lim_{\tau\rightarrow0}\dot{W}(x,\yvec \rightarrow x',\yvec';\tau)=\dot{W}(x,\yvec \rightarrow x',\yvec';0)$ from the Fokker-Planck  \Eref{corrFPE_all},
\begin{align}
\dot{W} (x,\yvec \rightarrow  x',\yvec';0)
= 
\Bigg\{
D_1\partial_{x'}^2 - u_1\partial_{x'}
 + \Nsum \left(
D_2\partial_{y_i'}^2 -u_2\partial_{y_i'}
+V_1'(x-y_i)\partial_{x'}
+V_2'(y_i-x)\partial_{y_i'}
\right)
\Bigg\}
\delta(x'-x)
\delta(\yvec'-\yvec) \ ,
\elabel{many_particles_transition_rate}
\end{align}
and the short-time limit of the logarithm of the ratio of 
propagators,
\begin{align}
    \lim_{\tau\rightarrow 0} \ln \left(
    \frac
    { W(x,\yvec \rightarrow x',\yvec';\tau)}
    { W(x',\yvec' \rightarrow x,\yvec;\tau)}
    \right)
    =&\Nsum\left\{\frac{x'-x}{2D_1}
    \left(\frac{u_1}{N}-V'_1(x-y_i)+\frac{u_1}{N}-V'_1(x'-y_i')\right)
    \right.
    \nonumber
    \\&
    \left.
    +\frac{y_i'-y_i}{2D_2}\left(u_2-V'_2(y_i-x)+u_2-V'_2(y_i'-x')\right)\right\} \ .
    \elabel{log_term}
\end{align}
Substituting \Erefs{many_particles_transition_rate}
and \eref{log_term} into \eref{stationary_entropy_production}, 
using the normalisation that accounts for B particles being
indistinguishable
$
\int_{0}^L\dint{x}\Ndint{y}P(x,\yvec)=N!
$, and 
integrating by parts, we obtain
\begin{align}
    \ep=
        \frac{1}{N!}
    \int_0^L\dint{x}\Ndint{y}\bigg\{
    &-\Nsum V''_1(x-y_i)
    +\frac{1}{D_1}\left(u_1-\Nsum V'_1(x-y_i)\right)^2
    \nonumber\\&
    -\Nsum V''_2(y_i-x)+\Nsum\frac{1}{D_2} \left(u_2-V'_2(y_i-x)\right)^2\bigg\}P(x,\yvec) \ .
\end{align}
Using the following marginalisation property of the $n$-point correlation
function,
\begin{equation}
    \int\dint{y_{n-1}} P^{(1+N)}_n(x,y_1,\ldots,y_{n-1})
    =(N-n+2) P^{(1+N)}_{n-1}(x,y_1,\ldots,y_{n-2}) \ ,
\end{equation}
we arrive at \Eref{ep_many_sheep_simplified}.
 
In the many-particle system with sinusoidal interaction potentials and zero drift,
the leading order 
behaviour of the entropy production is obtained by
substituting \Erefs{leading_order_two_point} and \eref{leading_order_three_point} into \Eref{ep_many_sheep_simplified}, 
\begin{align}
    \ep=\frac{2\pi^2 N }{L^2} \bigg[
    \bigg(\frac{a_1^2}{D_1}+\frac{a_2^2}{D_2}-\frac{(a_1+a_2)^2}{D_1+D_2}\bigg)
    \bigg(1-\frac{1}{8}\left(\frac{a_1+a_2}{D_1+D_2}\right)^2\bigg) 
    -(N-1)\frac{a_1(a_1+a_2)(D_1a_2-D_2a_1)^2}{D_1D_2(D_1+D_2)^2(4D_1+2D_2)}\bigg]
    + \hot \ ,
    \elabel{leading_order_ep}
\end{align}
shown in Fig.~\ref{fig:ep_chasing}.
In the validity regime, the leading 
order of the entropy
production in \eref{leading_order_ep} is 
nonnegative, as can be seen by
using the Cauchy-Schwarz inequality ${a_1^2}/{D_1}+{a_2^2}/{D_2}\geq {(a_1+a_2)^2}/{(D_1+D_2)}$.
In fact, the first term in the right-hand side of \Eref{leading_order_ep} vanishes if
$a_1/D_1=a_2/D_2$, which is, precisely,
the detailed balance condition for this case, \Eref{detailed_balance_two}.
However, outside of this validity regime,
\Eref{leading_order_ep} clearly provides an
unphysical result, as the entropy production rate
is predicted to be negative.
The remedy to this unphysical result lies
in the higher order corrections in the perturbation expansion.

\section{Doi-Peliti field theory of an interacting  particle system}
\label{sec:action_Fourier}
The Fokker-Planck equation of the joint probability density $P(x,\yset;t)$
of the system
described by \Eref{L_eqs} is
\begin{subequations}
\elabel{corrFPE_all}
\begin{align}
     \partial_t P(x,\yset;t)=&
\bigg\{D_1\partial^2_x-u_1\partial_x+\Nsum \left(D_2\partial^2_{y_i}-u_2\partial_{y_i}\right)\bigg\}P(x,\yset;t)
       \elabel{corrFPE0}
       \\&
       +\Nsum \bigg\{\partial_x \left(V_1'(x-y_i)P(x,\yset;t)\right)+\partial_{y_i}\left( V_2'(y_i-x)P(x,\yset;t)\right)\bigg\} \ ,
       \elabel{corrFPE}
\end{align}
\end{subequations}
where $P(x,\yset;t)$ is invariant under permutations of the positions $y_i$ due to the 
indistinguishability of B particles.

The Doi-Peliti action of this stochastic process 
follows from the Fokker-Planck \Eref{corrFPE_all}
\cite{TaeuberHowardVollmayr-Lee:2005,Cardy:2008,PruessnerGarcia-Millan:2022}.
We introduce the annihilation field $\phi(x,t)$ and the
Doi-shifted creation field $\tildephi(x,t)$
for species
A and, similarly, $\psi(y,t)$ and $\tildepsi(y,t)$
for species B. Even though there may be multiple 
particles of each species, the action functional $\AC[\phi,\tildephi,\psi,\tildepsi]$
is derived from 
the single particle dynamics \cite{PruessnerGarcia-Millan:2022}.
In this paper, we use the convention that an observable 
$\bullet$ is calculated via the path integral
$\ave{\bullet} = \int \AD[\phi,\tildephi,\psi,\tildepsi] \bullet\exp{\AC[\phi,\tildephi,\psi,\tildepsi]}$.
We split the action into two parts, 
$\AC=\AC_0+\actionPert$, where $\AC_0$, derived from 
\eref{corrFPE0}, governs
 free motion of particles,
\begin{equation}
    \AC_0
    =\int \dint t\int_0^L\dint x \tildephi\bigg\{-\partial_t+D_1\partial_x^2-u_1\partial_x\bigg\}\phi+ \int \dint t\int_0^L\dint {y} \tildepsi\bigg\{-\partial_t+D_2\partial_{y}^2-u_2\partial_{y}\bigg\}\psi
    \elabel{bare_action_two} \ ,
\end{equation}
and $\actionPert$, derived from \eref{corrFPE}, 
governs particle interactions
\cite{Doi:1976}
\begin{equation}
    \actionPert
    =-\int \dint t\int_0^L \dint x\int_0^L\dint {y}
    \left\{(\partial_x \tildephi)\phi\big[\partial_x V_1(x-y)\big](\tildepsi+1)\psi+(\partial_{y} \tildepsii)\psii\big[\partial_{y} V_2(y-x)\big](\tildephi+1)\phi \right\}
    \elabel{pert_action_two} \ .
\end{equation}
To derive $\AC_0$, we have used  that the number of particles is conserved  and, therefore,
$\int\dint x \tildephi(x,t)\phi(x,t)=1$
and $\int\dint{y} \tildepsi(y,t)\psi(y,t)=1$.
To derive the first term in $\actionPert$ we have used that the effect of interactions
on A particles at $x$ due to the presence of B particles
at $y$ is proportional to the number of B particles $y$, which is measured by the number  operator $\psi^\dagger\psi=(\tildepsi+1)\psi$, where $\psi^\dagger=(\tildepsi+1)$ is the creator field.
The second term in $\actionPert$ is derived similarly.
To calculate an observable $\bullet$, we do so order 
by order, in the perturbation expansion
$\ave{\bullet} = \ave{ \bullet\exp{\actionPert}}_0
=\ave{ \bullet\left({1+\actionPert}+\ldots\right)}_0$
about the Gaussian model given by $\AC_0$ assuming weak interactions.

We express the fields in Fourier space following the convention
\begin{equation}
    \ptwiddle{\phi}(x,t)=
     \frac{1}{L}\sum_n \int \dintbar{\omega}
     \exp{-\imag\omega t+\imag k_n x} \ptwiddle{\phi}_{n}(\omega) \
     \text{ and } \
    \ptwiddle{\phi}_{n}(\omega)=\int_0^L \dint y\int \dint t \exp{\imag\omega t-\imag k_n x}\ptwiddle{\phi}(x,t) \ ,
\end{equation}
where $\dbar=\dint/(2\pi)$ and $k_n=2\pi n/L$, and 
similarly for $\ptwiddle{\psi}$, as well as the interaction 
potentials, 
\begin{subequations}
\begin{align}
    V_1(x-y)&=\frac{1}{L}\sum_n \exp{\imag k_n (x-y)}\mu_n \ ,\\
    V_2(x-y)&=\frac{1}{L}\sum_n \exp{\imag k_n (x-y)}\nu_n \ .
\end{align}
\end{subequations}
The Gaussian part of the action in \Eref{bare_action_two} in Fourier
space reads
\begin{align}
    \AC_0 & \left[\phi_{n}(\omega), \phitilde_{n'}(\omega'), \psi_{n}(\omega), \psitilde_{n'}(\omega')\right]
    = \int \dintbar{\omega} \dintbar{\omega'} \deltabar(\omega+\omega')
    \frac{1}{L^2} \sum_{n,n'=-\infty}^\infty L\delta_{n,n'}
    \nonumber\\&\times \left\{ 
    \phitilde_{n'}(\omega')(-\imag\omega+D_1k_{n}^2-\imag u_1 k_n)
    \phi_{n}(\omega)
    + \psitilde_{n'}(\omega')(-\imag\omega+D_2k_{n}^2-\imag u_2 k_n)
   \psi_{n}(\omega)
    \right\} \ ,
\elabel{A0_Fourier}
\end{align}
and the part governing interactions in
\Eref{pert_action_two} is,
\begin{align}
    \actionPert &
    \left[\phi_{n}(\omega_1), \phitilde_{n'}(\omega_1'), \psi_{n}(\omega_2), \psitilde_{n'}(\omega_2')\right]\\
    =&
    \int\dintbar{\omega_1}\dintbar{\omega_1'}\dintbar{\omega_2}\dintbar{\omega_2'}
    \deltabar({\omega_1}+
    {\omega_1'}+{\omega_2}+{\omega_2'})
    \frac{1}{L^4}\sum_{n,n',m,m'}
    L\delta_{n+n'+m+m',0} 
    \nonumber\\ &\times
    \left(\mu_{-n-n'} k_{n'} k_{-n-n'}+
    \nu_{-m-m'} k_{m'} k_{-m-m'}\right)
    \phitilde_{n'}(\omega_1')\phi_{n}(\omega_1)\psitilde_{m'}(\omega_2')\psi_{m}(\omega_2)
    \nonumber\\
    +&
    \int\dintbar{\omega_1}\dintbar{\omega_1'}\dintbar{\omega_2}
    \deltabar({\omega_1}+{\omega_1'}+{\omega_2})
    \frac{1}{L^3}\sum_{n,n',m} L\delta_{n+n'+m,0} 
    \mu_{-n-n'} k_{n'} k_{-n-n'}
    \phitilde_{n'}(\omega_1')\phi_{n}(\omega_1)\psi_{m}(\omega_2)
    \nonumber\\ +&
    \int\dintbar{\omega_1}\dintbar{\omega_2}\dintbar{\omega_2'}
    \deltabar(
    {\omega_1}+{\omega_2}+{\omega_2'})
    \frac{1}{L^3}\sum_{n,m,m'}
    L\delta_{n+m+m',0} 
    \nu_{-m-m'} k_{m'} k_{-m-m'}
    \phi_{n}(\omega_1)\psitilde_{m'}(\omega_2')\psi_{m}(\omega_2) \ .
\end{align}
The bare propagators in Fourier space read, from \Eref{A0_Fourier},
\begin{subequations}
\elabel{bare_props}
\begin{align}
    \Abareprop{n,\omega}{n',\omega'}    \eqhat &
    \ave{\phi_n(\omega)\tildephi_{n'}(\omega')}_0= \frac{L\delta_{n+n',0}\deltabar(\omega+\omega')}{-\imag\omega+D_1 k_n^2+\imag u_1 k_n+r}
    \ ,\\
    \Bbareprop{n,\omega}{n',\omega'}  \eqhat &
    \ave{\psi_{n}(\omega)\tildepsi_{n'}(\omega')}_0= \frac{L\delta_{n+n',0}\deltabar(\omega+\omega')}{-\imag\omega+D_2 k_n^2+\imag u_2 k_n+r}
    \ ,
\end{align}
\end{subequations}
where $r$ is the mass,
$\deltabar=2\pi \delta$ is the Dirac $\delta$ function and $\delta_{n,m}$ is the Kronecker $\delta$ 
function. The mass $r$ is to be taken to zero after inverting the Fourier transform.
In \Eref{bare_props}, the diagrams 
\Abareprop{}{}
and
\Bbareprop{}{} represent the bare propagators of species A
and B respectively, where time is read from right to left.
The nonlinear couplings in $\actionPert$ involve a
four-point vertex 
and three-point vertices,
\begin{subequations}
\elabel{interaction_vertices}
\begin{align}
    \Avertex{n'}{n}{m'}{m}
    \eqhat & \frac{1}{L^3}
    \left(\mu_{-n-n'} k_{n'} k_{-n-n'}+
    \nu_{-m-m'} k_{m'} k_{-m-m'}\right)
    \delta_{n+n'+m+m',0}\deltabar(\omega_1+\omega_1'+\omega_2+\omega_2') 
    \ , \elabel{four_leg_vertex_A}\\
    \twoDthreelegA{n'}{n}{m} \eqhat &
   \frac{1}{L^2}\mu_{-n-n'}k_{n'} k_{-n-n'}\delta_{n+n'+m,0}\deltabar(\omega_1+\omega_1'+\omega_2)\elabel{three_leg_vertex_A} \ ,\\
     \twoDthreelegB{m'}{n}{m} \eqhat & \frac{1}{L^2}\nu_{-m-m'}k_{m'} k_{-m-m'}\delta_{n+m+m',0}\deltabar(\omega_1+\omega_2+\omega_2') \ .
\end{align}
\end{subequations}
The dashed line 
\tikz[baseline=-2.5pt]{\draw[Bactivity] (0,0) -- (-1,0);}
represents an interaction between two particles mediated by a potential. 
Diagrammatically, a potential has a similar 
representation as a propagator, with the difference
that it is represented vertically, since the 
potential has no time dependence
because its effect is \emph{propagated} instantly.

\section{Two-point correlation function}
\label{app_recurrence_higher_loops}

\subsection{Two-particle system}
The stationary  correlation function 
$P_2^{(2)}(x,y)=\lim_{t\to\infty}\ave{\phi(x,t)\psi(y,t)\phi^\dagger(x_0,t_0)\psi^\dagger(y_0,t_0)}$
gives the probability that particles A and B
are found at positions $x$ and $y$, respectively,
in the limit $t\to\infty$.
Since the system is finite and the dynamics are conservative,
the existence of the stationary limit is guaranteed.
Doi-shifting the creation fields in the observable, 
$\phi^\dagger=1+\tildephi$
and $\psi^\dagger=1+\tildepsi$, produces
$\ave{\phi\psi\phi^\dagger\psi^\dagger}=
\ave{\phi\psi\tildephi\tildepsi}
+\ave{\phi\psi}
+\ave{\phi\psi\tildephi}
+\ave{\phi\psi\tildepsi}
$, where the last three terms are all zero
because there is no matching vertex in $\actionPert$.

The stationary correlation function 
\begin{align}
P_2^{(2)}(x,y)=\lim_{t\rightarrow\infty}\ave{\phi(x,t)\psi(y,t)\tildephi(x_0,t_0)\tildepsi(y_0,t_0)}
\eqhat
\noloop
+ \zeroloop
+ \oneloop
+ \twoloop
+ \ldots 
\end{align}
is independent of the initial conditions $x_0$, $y_0$
since the only contributing Fourier modes in the limit $t\to\infty$ 
are the zeroth modes $\tildephi_0$ and $\tildepsi_0$
\cite{Zhen_2022, Garcia-MillanPruessner:2021}.
This is 
reflected in the diagrams by their
amputated incoming legs. 
We expand the Fourier space representation of the
two-point correlation 
\begin{equation}
    P_2^{(2)}(x,y) = \frac{1}{L^4}\sum_{n=-\infty}^\infty
    \exp{\imag k_n(x-y)} \ave{\phi_n\psi_{-n}\tildephi_0\tildepsi_0}
    \ ,
\end{equation}
perturbatively in small $\mu$ and 
$\nu$ compared to the diffusion constants $D_1$ and $D_2$,
\begin{equation}
    \ave{\phi_n\psi_{-n}\tildephi_0\tildepsi_0}
    = \sum_{j=0}^\infty \frac{1}{j!} \ave{\phi_n\psi_{-n}\tildephi_0\tildepsi_0 (\actionPert)^j }_0
    =\sum_{j=0}^\infty G_j(n) \ .
\end{equation}
The $j$-th order in the perturbation expansion of the $n$-th Fourier mode of $P_2^{(2)}$
is, diagrammatically,
\begin{equation}
    G_j(n) \eqhat \jminusoneloop \ ,
    \elabel{Gjn_diagram}
\end{equation}
where the incoming legs carry the zeroth Fourier mode only.
The $j$-th order correction $G_j$
is made of $j$ four-legged interaction vertices 
attached to $j$ propagators of each species
forming a chain of $j-1$ loops.
The $(j+1)$-th order  $G_{j+1}$ is generated from $G_{j}$ by attaching
an interaction vertex \eref{four_leg_vertex_A} to the outgoing legs,
effectively creating a new loop,
\begin{align}
    G_{j+1}(n) \eqhat & \sum_m \int\dintbar{\omega} \jloop \nonumber\\
    \eqhat & \frac{1}{L} \sum_m \int\dintbar{\omega} I_n
    \frac{\mu_{n-m} k_{-n}k_{n-m} + \nu_{-n+m}k_nk_{-n+m}}{
    (-\imag\omega+ D_1k_n^2+\imag u_1k_n)(\imag\omega+ D_2k_n^2-\imag u_2k_n)} G_{j}(m) \nonumber\\
    =& \frac{I_n}{L} \sum_m 
    \frac{(\mu_{n-m}+ \nu_{-n+m})k_{-n+m}}{
    (D_1+D_2)k_n+\imag (u_1-u_2)} G_{j}(m) \ ,
\elabel{app_recurrence_higher_loops}
\end{align}
where $I_n=1-\delta_{n,0}$ ensures that $G_{j}(0)=0$ for $j>0$.
The zeroth-order term is, trivially, the constant zeroth Fourier
mode $G_0(n)=L^2 \delta_{n,0}$.

The iterative scheme in \Eref{app_recurrence_higher_loops} has the 
same structure as the recurrence 
relation found in the stationary probability 
distribution of a drift-diffusive particle in
a periodic potential \cite{FehertoiPolackova:ToBePublished}.
Therefore, by mapping the two systems, the two-point correlation
function follows straight-forwardly, 
$P_2^{(2)}(x,y)=\rho(x-y)/L$, with
\begin{equation}
    \rho(\ell)= 
    \frac{1}{L \norm}
    \exp{-\frac{V_1(\ell)+V_2(-\ell)-(u_1-u_2)\ell}{D_1+D_2}}
    \int_{\ell}^{\ell+L}\dint{\ell'} 
    \exp{\frac{V_1(\ell')+V_2(-\ell')-(u_1-u_2)\ell'}{D_1+D_2}}
    \ ,
    \elabel{two_point_corr}
\end{equation}
where the constant $\norm$ is such that $\int_{0}^{L}\dint{\ell}\rho(\ell)=1$ \cite[Eq.~(11.37)]{Risken:1989}, shown in Fig.~\ref{fig:two_point_correlation}.
Although our argument is based on a 
perturbation expansion about small couplings $\nu$ and $\mu$, 
the exact result in \Eref{two_point_corr}
is valid for any couplings
because the map between $G_{j+1}$ in 
\Eref{app_recurrence_higher_loops} and 
the analogous $G_{j+1}$
in \cite{FehertoiPolackova:ToBePublished} 
holds for all orders $j\in\{0,1,\ldots\}$.

\subsection{Many-particle system}

In the many-particle system, the two-point correlation function $P_2^{(1+N)}(x,y)$
is the probability of finding the only particle 
of species A at position $x$, and one out of $N$ particles of 
species B at $y$.
This observable 
is represented as the following path integral and
diagrammatic perturbation expansion,
\begin{align}
    P_2^{(1+N)}(x,y)= & \lim_{t\to\infty}
    N
    \ave{\phi(x,t)\psi(y,t)
    \phi^\dagger(x_0,t_0)
    \psi^\dagger(y_1,t_0)
    \psi^\dagger(y_2,t_0)
    \ldots
    \psi^\dagger(y_N,t_0)} 
    \nonumber\\
         \eqhat &\lim_{t\rightarrow\infty}
          N
          \left(
\noloop \
+ \zeroloop \
+ \oneloop \
+ \twoloop \
+ \dots
           \right.
           \nonumber\\
           &\left. +(N-1)\left(
           \threeToTwoPointB \
           +\threeToTwoPointA \
           +\threeToTwoPointC \
           +\dots\right)+\dots\right)
           \ ,
           \elabel{app_two_point_diagrams}
     \end{align}
up to third order in the interaction vertices.
In principle, there are more terms that enter in this observable than those shown in \Eref{app_two_point_diagrams}.
However, terms such as
\begin{align}
\lim_{t\to\infty}
\threeToTwoPointE \
=
\lim_{t\to\infty}
\threeToTwoPointD \
=
\lim_{t\rightarrow\infty}
\threeToTwoPointC \
    \eqhat 0 
\end{align}
do not contribute in the limit $t\to\infty$ because of factors $k_0$
in the interaction vertices in \Eref{interaction_vertices}.

The sinusoidal interaction potentials $V_1(\ell)=a_1\cos{k_1 \ell}$ and $V_2(\ell)=a_2\cos{k_1 \ell}$ only have two
nonzero Fourier modes,
\begin{align}
    \mu_1=\mu_{-1}=\frac{a_1 L}{2} \ , && 
    \nu_1=\nu_{-1}=\frac{a_2 L}{2} \ , 
\end{align}
and $\mu_n=\nu_n=0$ for $n\neq\pm1$,
and we further assume zero drift $u_1=u_2=0$.
The terms in \Eref{app_two_point_diagrams} involving 
the only A particle and one of the B particles are, therefore,
\begin{subequations}
\elabel{two_point_contributions_1A1B}
\begin{align}
    \lim_{t\rightarrow\infty}
    \noloop
    &\eqhat  \frac{1}{L^4}\sum_n\exp{\imag k_n(x-y)}G_0(n)=
    \frac{1}{L^2}\\
   \lim_{t\rightarrow\infty}
   \zeroloop
   &\eqhat\frac{1}{L^4}\sum_n\exp{\imag k_n(x-y)}G_1(n)
   =-\frac{1}{L^2}\frac{a_1+a_2}{D_1+D_2}\cos k_1(x-y)\\
    \lim_{t\rightarrow\infty}
    \oneloop
    &\eqhat\frac{1}{L^4}\sum_n\exp{\imag k_n(x-y)}G_2(n)
    =\frac{1}{L^2}\left(\frac{a_1+a_2}{D_1+D_2}\right)^2\frac{\cos{k_2(x-y)}}{4}\\
     \lim_{t\rightarrow\infty}
     \twoloop
     &\eqhat\frac{1}{L^4}\sum_n\exp{\imag k_n(x-y)}G_3(n)
     =\frac{1}{L^2}\bigg(\frac{a_1+a_2}{D_1+D_2}\bigg)^3
     \left(\frac{\cos{k_1 (x-y)}}{8}-\frac{\cos{k_3 (x-y)}}{24}\right)\ ,
\end{align}
\end{subequations}
where $G_j(n)$ is defined in \Eref{Gjn_diagram}.
The terms in \Eref{app_two_point_diagrams}
involving two B particles are,
\begin{subequations}
\elabel{two_point_contributions_1A2B}
\begin{align}
    \lim_{t\rightarrow\infty} \threeToTwoPointB \
   \eqhat  &
   \frac{1}{L^5} \sum_{n,i} \exp{\imag k_n (x-y)} \int\dintbar{\omega}\dintbar{\omega'} 
\nonumber   \\&
   \frac{(\mu_n k_{-n}k_n + \nu_{-n}k_n k_{-n}) \mu_{-i} k_{-n} k_{-i} (\mu_i k_{-n-i}k_i+\nu_{-i}k_i k_{-i})}{(-\imag\omega+D_1 k_n^2+r)(-\imag\omega +D_1k_n^2+r)(\imag\omega+D_2k_n^2+r)(-\imag(\omega+\omega')+D_1k_{n+i}^2+r)(\imag\omega'+D_2k_i^2+r)}
\nonumber   \\
    =&\frac{1}{L^2}\frac{a_1(a_1+a_2)(a_1D_2-a_2D_1)}{4D_2(D_1+D_2)^2(2D_1+D_2)}{\cos{k_1(x-y)}}     \\
    \lim_{t\rightarrow\infty}\threeToTwoPointA \
    \eqhat &
    \frac{1}{L^5}\sum_{n,i}\exp{\imag k_n(x-y)} \int\dintbar{\omega}\dintbar{\omega'}
\nonumber    \\&
    \frac{(\mu_{i}k_{-i}k_{i}+\nu_{-i}k_{i}k_{-i})\mu_{-i}k_{-n}k_{-i} (\mu_n k_{-n-i}k_n+\nu_{-n}k_nk_{-n}) }{(-\imag\omega+D_1k_n^2+r)(-\imag(\omega+\omega')+D_1k_{n+i}^2+r)(-\imag\omega'+D_1k_i^2+r)(\imag\omega'+D_2k_i^2+r)(\imag\omega+D_2k_n^2+r)}
\nonumber    \\
    =&\frac{1}{L^2}\frac{a_1(a_1+a_2)(a_1D_2-a_2D_1)}{4D_2(D_1+D_2)^2(2D_1+D_2)}{\cos{k_1(x-y)}} \ .
\end{align}
\end{subequations}
Using \Erefs{two_point_contributions_1A1B}
and \eref{two_point_contributions_1A2B} in 
\Eref{app_two_point_diagrams}, we obtain
the two-point correlation function perturbatively up to third order,
\begin{align}
    P^{(1+N)}_2(x,y)=&\frac{N}{L^2}\bigg[1-\frac{a_1+a_2}{D_1+D_2}\cos{k_1 (x-y)}+\frac{1}{4}\bigg(\frac{a_1+a_2}{D_1+D_2}\bigg)^2{\cos{k_2 (x-y)}} 
    \nonumber\\
    &
    +\frac{1}{24}\bigg(\frac{a_1+a_2}{D_1+D_2}\bigg)^3
    \left(3\cos{k_1 (x-y)}-\cos{k_3 (x-y)}\right) 
    \nonumber\\
    &
    +(N-1)
    \frac{a_1(a_1+a_2)(a_1D_2-a_2D_1)}{2D_2(D_1+D_2)^2(2D_1+D_2)}{\cos{k_1(x-y)}}
    \bigg]
    +\hot
    \ .
\elabel{leading_order_two_point}
\end{align}

\section{Three-point correlation function}
\label{sec:app-many-particles3}

The three-point correlation function of the position $x$ of the A particle and positions $y$ and $y'$ of two 
B particles in the stationary state is
\begin{align}
&     P_3^{(1+N)}(x,y,y')=  \lim_{t\to\infty}
    N(N-1)
    \ave{\phi(x,t)\psi(y,t)\psi(y',t)
    \phi^\dagger(x_0,t_0)
    \psi^\dagger(y_1,t_0)
    \psi^\dagger(y_2,t_0)
    \ldots
    \psi^\dagger(y_N,t_0)
    } \nonumber\\
       & \eqhat \lim_{t\rightarrow\infty} N(N-1)
        \left(
\threebareprop \
+ \threepropA \
+ \threepropB \
+ \threepropAA \
+ \threepropBB \
+ \threepropAB \
+ \threepropBA \
+ \dots\right) \ .
\elabel{P3_obs}
\end{align}
We assume sinusoidal interaction potentials $V_1(\ell)=a_1\cos{k_1 \ell}$ and $V_2(\ell)=a_2\cos{k_1 \ell}$, 
and zero drifts $u_1=u_2=0$.
We distinguish two types of terms in \Eref{P3_obs}:
those where a propagator
\Bbareprop{}{} is 
decoupled from the other two propagators, and those
where all three propagators are coupled through 
interaction vertices. In the first case,
the corresponding diagrams can be calculated straightforwardly using the results in \eref{two_point_contributions_1A1B}
multiplying by $1/L$. For example,
\begin{equation}
\elabel{example}
    \threepropA \
    =\frac{1}{L} \zeroloop \
    \eqhat
    -\frac{1}{L^3}\frac{a_1+a_2}{D_1+D_2}\cos k_1(x-y) \ .
\end{equation}
Then, we only need to evaluate the
last two diagrams in \eref{P3_obs}, which
are symmetric under swapping $y$ and $y'$,
\begin{align}
\lim_{t\rightarrow\infty}\threepropAB \
 \eqhat &
 \frac{1}{L^5} \sum_{n,m}
 \exp{\imag k_n(x-y)+\imag k_m(x-y')}
 \int\dintbar{\omega}\dintbar{\omega'} 
 \nonumber \\&
 \frac{(\mu_mk_{-n-m}k_m+\nu_{-m}k_m k_{-m})(\mu_nk_{-n}k_n+\nu_{-n}k_nk_{-n})}{(-\imag(\omega+\omega')+D_1k_{n+m}^2+r)(-\imag\omega+D_1k_n^2+r)(\imag\omega+D_2k_n^2+r)(\imag\omega'+D_2k_m^2+r)} \nonumber\\
=&\frac{1}{2L^3}\frac{a_1+a_2}{D_1+D_2}   \bigg\{
 \frac{a_2}{2D_2}\cos{k_1(y-y')}+\frac{a_2+2a_1}{4D_1+2D_2}\cos{k_1(2x-y-y')}\bigg\}
 \ .
\elabel{diagram_xyy'}
 \end{align}
In fact, \Eref{diagram_xyy'} is invariant under exchange of
$y$ and $y'$, so the two last terms in the right-hand side of \eref{P3_obs} are therefore equal.
We obtain the 
three-point correlation function up to second order in
the perturbation expansion
by using \Erefs{two_point_contributions_1A1B} 
and \eref{diagram_xyy'}, 
 in \Eref{P3_obs},
\begin{align}
P^{(1+N)}_3(x,y,y')=&\frac{N(N-1)}{
L^3}\bigg[1-\frac{a_1+a_2}{D_1+D_2}(\cos{k_1 (x-y)}+\cos{k_1 (x-y')}) 
\nonumber \\
&
 +\frac{1}{4}\bigg(\frac{a_1+a_2}{D_1+D_2}\bigg)^2({\cos{k_2 (x-y)}}+
    {\cos{k_2 (x-y')}}) 
    \nonumber \\
& + \frac{a_1+a_2}{2(D_1+D_2)}   \bigg\{
 \frac{a_2}{D_2}\cos{k_1(y-y')}+\frac{2a_1+a_2}{2D_1+D_2}\cos{k_1(2x-y-y')}\bigg\}\bigg]
 +\hot
  \ .
\elabel{leading_order_three_point}
 \end{align}
The 
correlation functions in 
\Erefs{leading_order_two_point} and 
\eref{leading_order_three_point} can also be
obtained by performing a Taylor expansion 
about the equilibrium distribution in \Eref{DB_correlation},
with $a_1/D_1 = a_2/D_2$, where the normalisation
constant also needs to be expanded.

Marginalising \Eref{leading_order_three_point}, we obtain the
two-point correlation function between
two particles of species B at positions $y$ and $y'$, 
\begin{equation}
\elabel{sheep_sheep_corr}
    P_{B,B}^{(1+N)}(y,y')=\int_0^L\dint{x}P^{(1+N)}_3(x,y,y')=
    \frac{N(N-1)}{L^2}\left[1+\frac{a_2(a_1+a_2)}{2D_2(D_1+D_2)}\cos k_1(y-y')\right]+ \hot
    \ ,
\end{equation}
whose second-order correction indicates the effective interactions between particles of species B,
shown in Fig.~\ref{fig:sheep_sheep_corr}.

\end{document}